\documentstyle[12pt]{article}

\def\be{\begin{equation}}
\def\ee{\end{equation}}
\def\bea{\begin{eqnarray}}
\def\eea{\end{eqnarray}}

\def \phi{\varphi}
\def\slash#1{#1\!\!\!/\!\,\,}

\def\SMo{Standard Model}
\def\SM{\SMo\ } \def\SMp{\SMo.\ } 
 
\def\GBo{Goldstone Boson}
\def\GB{\GBo\ }  \def\GBs{\GBo s\ } \def\GBsp{\GBo s.\ }

\def\beq{\begin{equation}}      \def\eeq{\end{equation}}
\def\bea{\begin{eqnarray}}      \def\eea{\end{eqnarray}}
\def\bq{\begin{quote}}          \def\eq{\end{quote}}

\begin{document}

 \begin{flushright}
 TUM--HEP--270/97
 \end{flushright}

\

\centerline{\large\bf Phenomenologically Viable 
 Dynamical Electro--Weak Symmetry Breaking}

\

\

\centerline{Manfred Lindner}

\

\begin{center}
{\sl Physik Department, Technische Universit\"at M\"unchen\\
James--Franck Str., D--85748 Garching b. M\"unchen, Germany}
\end{center}

\

\

\centerline{\bf Abstract}

\

The lack of deviations from the Standard Model at the current
level of experimental precision can be explained systematically 
in suitable models of dynamical electro--weak symmetry breaking. 
The key ingredient is dynamics which produces a scalar unitarity 
partner for the Goldstone--Bosons and which leads to a 
decoupling of effects beyond the Standard Model. 
A phenomenologically successful left--right symmetric model 
is presented from this point of view.

\ \vskip 5.5cm

\hfill

\hrule

{\small To appear in the Proceedings of the Workshop on The 
Higgs puzzle-- What can we learn from LEP II, LHC, NLC and
FMC?, Ringberg Castle, Germany, December 8-13, 1996. Ed. B. Kniehl, 
World Scientific, Singapore}

\clearpage

The current status of elementary particle physics is remarkable: 
The \SM describes all existing data with impressive precision 
including many different reactions and impressive precision 
tests of radiative corrections \cite{SMreview}. There is however 
consensus among theorists that there should be physics beyond 
the \SM in the $TeV$ regime related to the solution of the 
famous hierarchy problem \cite{HP}. The phenomenological discussion of 
physics beyond the \SM is these days however often replaced by 
a comparison of two simple representatives of the two 
main directions: The Minimal Supersymmetric Standard 
Model (MSSM) as simplest realization of supersymmetry (SUSY)
and Technicolor (TC) as a simple dynamical symmetry breaking
(DSB) scenario. In comparisons it is often stressed 
how good the MSSM works, and how difficult naive TC is in the 
light of the existing precision data.
This is sometimes even used to suggest that there is already a 
hint towards the solution of the hierarchy problem in the data. 
The aim of this article is to stress that this is misleading.
The comparison should not be viewed as a hint for or against 
either supersymmetry or dynamical symmetry breaking in general. 
The current situation is simply the success of {\bf any} 
model that can ``hide'' behind the \SMp In other words, it is 
a consequence of theories, where new effects decouple when 
physics beyond the \SM is made very heavy. We will point out
that it is possible to build systematically models of 
dynamical symmetry breaking with such a decoupling limit. 

The MSSM and most other SUSY models have a decoupling limit. 
If all particles beyond the \SM are sent to the
$TeV$ scale then one obtains effectively the \SM plus deviations
which are today not accessible by experiments. There is however 
a SUSY relic which stems from the fact that the Higgs mass 
is at tree level below the $Z$ mass. Including radiative 
corrections proportional to $m_t^4$ the Higgs mass can be somewhat 
larger up to $m_H\simeq 130~GeV$. Non--minimal SUSY models 
have similar decoupling limits and (for different reasons) 
a Higgs mass range \cite{SUSYmH}  $m_H \leq 150~GeV$.

Naive Technicolor (TC), on the other side is based on rescaled 
QCD with a more or less (up to large $N_{TC}$ corrections) 
fixed spectrum of bound states. The Techni--pions, the \GBs 
which give mass to $W$ and $Z$ in a dynamical Higgs mechanism, 
do not have a scalar sigma--like particle which acts as unitarity 
partner. Instead low lying composite vector resonances, the 
Techni--rhos, play the role of the unitarity partner. This is 
a consequence of QCD--like dynamics which does not produce a 
Higgs--like scalar and which does not have a decoupling limit. 
The Techni--rho states lead however to severe phenomenological 
problems: Being vector particles they can mix with the $W$ and $Z$ 
leading thus for typical masses to unacceptable effects in the 
precision variable $S$. This problem becomes even more severe 
as $N_{TC}$ is increased, since for large $N_{TC}$ the 
Techni--rho to Goldstone Boson mass ratio becomes smaller, 
and the mixing with $W$ and $Z$ is increased. Most TC scenarios 
have additionally extra Techni--fermion doublets which must be 
heavy. The mechanism responsible for the top--quark to 
bottom--quark mass ratio will unavoidably also induce a mass 
splitting in the Techni--fermion doublet. This leads to 
completely unacceptable effects in terms of the precision 
variable $T$ which measures custodial $SU(2)$ violation.

Note, however, that neither the success of the MSSM nor the 
problems of naive TC stem generically from a supersymmetric or 
dynamical solution of the hierarchy problem. The phenomenological 
viability or failure is to a large extent simply the presence 
or absence of the mentioned decoupling limit and the lack of 
deviations of the experimental data from the \SMp This is the 
success of models with scalar Higgs particles over models with 
low lying vector particles and the fact whether the 
scalars or vectors are composite or fundamental does not play 
a role in this consideration. It is therefore tempting to ask 
if it is possible to build systematically viable models of 
dynamical symmetry breaking with a decoupling limit. Such 
models should contain an effective or composite Higgs particle 
instead of vector resonances and would therefore look effectively 
much like the \SMp Deviations from the \SM (if e.g. necessary 
in quantities like $R_b$ etc.) would be connected to couplings 
and particles beyond the \SM and might be explained by lowering 
some of the additional states. Due to the underlying dynamics 
it should however not be possible to obtain completely arbitrary 
\SM parameters. Mass restrictions for the composite Higgs mass 
as a relic of the underlying scenario are therefore expected.

The existence of DSB models with a such a decoupling limit is 
an important question 
since it would lead systematically to viable models of 
dynamical electro--weak symmetry breaking. This is of course 
directly related to the solution of the hierarchy problem and 
therefore to the absence of huge quadratic corrections in the 
scalar symmetry breaking sector. In supersymmetry huge 
quadratic corrections are cancelled due to restored supersymmetry 
above the SUSY breaking scale $\Delta\simeq TeV$. Dynamical 
solutions of the problem on the other side provide a form--factor 
(i.e. unbinding) in the scalar sector, again at scales 
${\cal{O}}(TeV)$. The difference is therefore how huge quadratic 
corrections are avoided beyond $TeV$ scales: Are they dissolved 
due to the unbinding of some dynamics or has nature arranged 
a huge systematic cancellation due 
to SUSY? Both routes have attractive features and experiments 
must ultimately find out which direction is chosen by nature. 
There are however genuine technical differences between SUSY and
DSB which must be kept in mind. Such technical differences 
should not be used as argument for or against one of the 
directions since technicalities are just human problems
and not conceptual problems. The point is that SUSY is essentially 
perturbative physics, while DSB is generically non--perturbative. 
Many quantities in DSB are therefore hard to calculate and the 
answers are often very rough. In 
DSB there is also no or little guidance from a greater picture 
like in more ambitious SUSY scenarios. Note also that among the 
arguments for SUSY there is no argument for ``SUSY now'', i.e. 
SUSY at the electro--weak scale. There could easily be 
non--supersymmetric physics at the $TeV$ scale responsible for
electro--weak symmetry breaking which is embedded 
into a greater supersymmetric picture which takes us to the GUT 
or Planck scale. 
DSB models have also number of attractive features. Most important 
is the fact that DSB is a mechanism which is always possible. It must 
not be artificially invented, and it occurs automatically in  
quantum field theory if the energy of the ground state can be 
lowered in this way. Symmetry breaking is a simple and natural side effect.
DSB is known to be realized in many systems in other areas of physics. 
Well known examples are ferro--magnetism 
and superconductivity, where the latter is even an example for a 
dynamical Higgs mechanism. 

It is important to distinguish between the Higgs mechanism 
and a Higgs particle. The Higgs mechanism of the \SM requires only 
an operator $\hat O$ with some general properties:
\begin{itemize}
\item 
Lorentz invariance of the vacuum requires that $\hat O$ must 
be scalar
\item Symmetry breaking requires $\langle \hat O\rangle \neq 0$
\item 
$\hat O$ must transform as a doublet of $SU(2)_L$ with hypercharge 
$Y=1$
\item 
$\hat O$ must however not necessarily be fundamental, i.e. a 
scalar field.
\end{itemize}
For $\langle \hat O\rangle \neq 0$ (i.e. in unitary gauge)
one can write $\hat O = \left(\langle \hat O\rangle + 
\delta \hat O\right) e^{i\phi_a T_a}$ which results
in a dynamical Higgs mechanism where the \GBs $\phi_a$ are
eaten if a $|D\hat O|^2$ term is present. $\delta \hat O$ 
corresponds to the spectrum of the theory which may or may 
not contain a composite scalar Higgs particle $H$.
If a scalar exists, then there is also an interaction 
potential $V(H)$ which reduces with the additional 
requirement of renormalizability to $\lambda \Phi^4$.
Note, however, that the existence of a scalar, of a potential 
and renormalizability are in principle not necessary if the 
Higgs mechanism is only effective. Only the underlying theory 
should be a renormalizable! The scalar \GB spectrum and 
the Higgs mechanism follow however already from the involved global 
symmetries independently of the nature of $\hat O$. 
Many properties of the \GBs and their interactions can be 
understood in terms of symmetries and the details of the 
interaction responsible for symmetry breaking are therefore
almost irrelevant. This is well established in QCD, where 
even a Nambu--Jona-Lasinio (NJL) description of chiral
symmetry breaking (which has almost nothing to do with QCD,
but breaks the chiral symmetries correctly) leads 
due to the corresponding Ward--Identities to 
remarkable good description of pion interactions.

A scalar Higgs particle has so far not been seen
and until recently the excitations $\delta \hat O$ were
essentially unconstraint. The now existing precision data 
disfavour however a spectrum with low lying vectors (like in TC)
and favour a spectrum were a Higgs--like scalar plays the 
role of the unitarity partner of the \GBsp  Such a spectrum, 
i.e. the existence of scalar Higgs (bound) states is in DSB clearly 
a dynamical issue. One should therefore try to build models 
with composite scalars along this line 
which also naturally avoid problems with the phenomenological
points below. If this is possible, then 
the resulting models should be phenomenologically much more 
viable than e.g. naive TC. The conceptually challenging point 
is to understand which dynamics leads naturally to such 
models.

The known experimental features should be used in bottom up
approaches of DSB models of electro--weak symmetry breaking.
In other words only ingredients which do not lead to models 
which are grossly wrong should be used. The fact that the known 
particles, reactions and especially radiative corrections 
(e.g. in terms of the observables $S$, $T$ and $U$) agree 
very well with the \SM leads to the following considerations:

\begin{itemize}
\item 
$T\simeq m_t^2$ agrees very good with the \SM value. As mentioned
above this disfavours scenarios which have sizable extra 
custodial $SU(2)$ violating effects. Thus extra fermionic 
doublets beyond the \SM should be avoided.
\item
$S$ agrees very well with the \SM value. This means that radiative
corrections from scalar loops are acceptable, but mixings with
strongly coupled vector states ${\cal{O}}(TeV)$ like the Techni--rho 
lead to unacceptable large contributions. The underlying dynamics 
should therefore {\bf not} be scaled QCD with low lying composite 
vectors, but another dynamics which produces systematically a scalar 
spectrum. 
\item
The absence of flavour changing neutral currents (FCNC's) beyond the
\SM is a severe problem in models where the top mass is explained
after the electro--weak symmetry is broken. The generation of quark 
masses in extended technicolor is a well known example. The situation 
is however much better if the top quark is directly involved in the 
electro--weak symmetry breaking. Thus a top--condensate should play 
an essential role.
\end{itemize}

If it is possible to build models along this line with a decoupling
limit then one should obtain systematically models which are 
phenomenologically much more viable than e.g. naive TC. It should 
be obvious that the necessary ingredients 
are not very restrictive or artificial. 
The physically most interesting point is probably to understand 
which sort of dynamics produces systematically a scalar spectrum.

There is a nice prototype model which realizes the required features 
in an excellent and minimalistic way: The so--called BHL model \cite{BHL} 
of electro--weak symmetry breaking. The idea is to eliminate in the
\SM the fundamental Higgs field and to introduce instead a new 
attractive interaction which can lead to a suitable scalar operator 
$\hat O$ which can condense. The heaviness of the top quark and the 
need to explain its mass together with electro--weak symmetry breaking 
(to avoid FCNC problems) suggests that $\hat O \simeq \bar Lt_R$ 
composed of the right handed top field and the left--handed doublet 
$L=(t_L,b_L)^T$. The simplest attractive interaction that can lead 
to the desired condensate is a four--fermi term. One arrives therefore 
at the Lagrangian 
\beq
{\cal L} = {\cal L}_{kin}(g,f) + G~\bar L t_R \bar t_R L~,
\label{L4f}
\eeq
where ${\cal L}_{kin}(g,f)$ represents the kinetic terms
for the known gauge and chiral fermion fields. The 
above Lagrangian represents a gauged NJL--model, where 
condensation and electro--weak symmetry breaking can occur
for $G>G_{critical}$.
In auxiliary field formalism one can define the local
composite operator $\phi := -G\bar t_RL$ which allows to 
rewrite the Lagrangian (\ref{L4f}) by means of the equations 
of motion into
\beq
{\cal L} = {\cal L}_{kin}(g,f) 
- \bar L \phi t_R - \bar t_R\phi^+L -G^{-1}\phi^+\phi~.
\label{Laux}
\eeq
Loop effects generate (for $\Lambda \gg v$) all the missing 
renormalizable terms\footnote{For $\Lambda \simeq v$ many
non--renormalizable terms would be generated in addition}. 
Including quantum effects one arrives thus at 
\bea
{\cal L}_{eff} & = &
{\cal L}_{kin}(g,f) + \delta{\cal L}_{kin}(g,f)\\
& & + Z_\phi |D\phi|^2 
- (1+\delta g_t) \left(\bar L \phi t_R + \bar t_R\phi^+L\right) \\
& & - (G^{-1} - \delta M^2)\phi^+\phi 
- \frac{\delta\lambda}{2}(\phi^+\phi)^2 ~.
\label{Leff}
\eea
This is the effective action for the composite operator 
$\hat O \equiv \phi$. We can read off the CJT effective 
potential \cite{CJT} in fermion the bubble approximation
\beq
V_{eff}(\phi)=(G^{-1} - \delta M^2)\phi^+\phi 
+ \frac{\delta\lambda}{2}(\phi^+\phi)^2 ~.
\label{Veff}
\eeq
The terms $\delta M^2$ and $\delta \lambda$ follow from
the one loop diagrams with two and four external composite
operators connected via a vertex with a factor $G^{-1}$.
The potential (\ref{Veff}) leads to symmetry breaking for 
small $G^{-1}$, i.e. large enough $G>G_{critical}$. 
Up to the unconventionally normalized kinetic term in 
the effective Lagrangian (\ref{Leff}) everything looks 
like the Higgs sector of the \SMp  The difference is the
the fact that the Higgs is composite which leads to parameter 
restrictions. If the infrared cutoff $\mu$ is increased 
towards $\Lambda$ then the quantum effects which generate 
the full Lagrangian must disappear. From this one obtains 
the so--called ``compositeness conditions'' 
\beq
Z_\phi 
\stackrel{\mu^2\rightarrow\Lambda^2}{\longrightarrow} 0~;
\quad
\delta M^2
\stackrel{\mu^2\rightarrow\Lambda^2}{\longrightarrow} 0~;
\quad
\delta\lambda
\stackrel{\mu^2\rightarrow\Lambda^2}{\longrightarrow} 0~.
\label{cond}
\eeq
The simple rescaling $\phi\longrightarrow\phi/\sqrt{Z_\phi}$ 
finally leads to a kinetic term for the composite Higgs field 
which is normalized to unity. If the compositeness conditions 
are rewritten in this way then one obtains boundary conditions 
which the \SM must fulfil if the Higgs stems from top condensation.
These boundary conditions can be imposed on the renormalization 
group flow (RGE) of the \SM and one obtains rather stable 
predictions due to the so--called infrared quasi 
fixed--points \cite{IRqFP}. The resulting top mass prediction 
is however even for undesirable large scales $10^{15}~GeV$ 
about 30 percent too high. For more desirable scales of 
new physics in the multi $TeV$ regime the top mass is 
too high by about a factor two and the BHL model is therefore 
phenomenologically unacceptable.
The BHL model demonstrates nevertheless nicely 
how the decoupling of physics beyond the \SM can be realized.

The four--fermion interactions of the BHL model are however
undesirable for fundamental physics and it is tempting to ask 
which physics can be described effectively in this way. The 
four--fermion term of the model changes by a Fierz transformation 
into the remarkable simple current--current structure:
\beq
G\bar{L}t_R\bar{t_R}L 
\stackrel{Fierz}{\longrightarrow}
-\frac{G}{2} \left(\bar{L}\gamma_\mu L\right)
\left(\bar{t}_R\gamma_\mu t_R\right)~.
\label{fierz}
\eeq
It is therefore tempting to relate the four--fermion structure 
of the BHL model to the exchange of suitable massive, strongly 
coupled vector bosons. Due to the success of renormalizable gauge 
theories one would therefore be tempted to justify the whole 
scenario by a broken, extended gauge group where a massive
boson propagator has been integrated out. This idea has been realized 
and a number models along this line have been 
proposed \cite{topcolor,topcolor2,U1,SU2V}. Note that the dynamics 
of such scenarios deviates clearly from QCD in this picture.
There are even hints for interesting confinement--Higgs
dualities which might play a role in such models \cite{topcolor2}.

All these attempts have however the problem that they 
produce a top mass which is unacceptably high. This is not an 
accident, but has a systematic reason. For an asymptotically 
free theory the dynamically generated top propagator can be 
written as
\beq
S_t = \frac{i}{\slash{p}-\Sigma_t(p^2)} ~,
\label{St}
\eeq
where 
\beq
\Sigma(p^2)\stackrel{p^2\rightarrow\infty}{\longrightarrow}0~.
\label{ass}
\eeq
The so--called Pagels--Stokar relations \cite{PS} which are mostly 
based on Ward--Identities relate the dynamically generated propagator
to the associated Goldstone Boson decay constant by
\beq
F_\pm^2 = \frac{N_c}{32\pi^2}
\int dk^2 k^2 \frac{\Sigma_t(k^2)}{(k^2-\Sigma_t(k^2))k^2}~.
\label{PS}
\eeq
Note that the above integral is formally log divergent, but
finite if the asymptotic behaviour eq.~(\ref{ass}) is taken into 
account. The Pagels--Stokar relation is very useful, since it 
is a powerful relation between the dynamically generated top mass, 
the Goldstone Boson decay constant $F_\pm$ and $m_W^2=g_2^2 F_\pm^2$.
It is instructive to observe that the integral on the {\it rhs} of the 
Pagels--Stokar relation, eq.~(\ref{PS}), feels the structure of 
$\Sigma_t$ only on a logarithmic scale. Without a specific theory 
one could for example use the simple approximation 
\beq
\Sigma_t(p^2) = m_t \Theta(\Lambda^2-p^2)~.
\label{ansatz}
\eeq
Inserting this into eq.~(\ref{PS}) and solving for the top mass 
results in
\beq
m_t^2 = \frac{32\pi^2 m_W^2}{N_c~g_2^2~\ln(\Lambda^2/m_t^2)}~,
\label{PSmt}
\eeq
which is exactly the relation which was obtained in the BHL model in
bubble approximation. This makes sense, since the ansatz (\ref{ansatz})
corresponds to a Nambu--Jona-Lasinio (NJL) gap equation. Corrections 
to this relation come like in the BHL model from other, weak gauge 
contributions and are expected to be moderate. The Pagels--Stokar
relation is mostly based on symmetries and it explains why most 
variants of the BHL model, 
like two Higgs doublets or the supersymmetric version, produce 
essentially a similar top mass which is too high.
One can conclude from this that dynamically electro--weak symmetry breaking 
can not be driven by a top condensate alone. In order to get acceptable
top mass values there must be at least a second condensate implying 
a more involved dynamical scenario. Such a second condensate could 
be related to further global symmetry breakings. If one beliefs in a 
continuation of the success story of renormalizable gauge theories
then it would even be natural to relate the extra condensate to 
the breaking of an extended gauge sector.
 
A class of models where this idea can be realized is given by 
dynamically broken left--right symmetric theories. 
Left--right symmetric models based on the gauge group 
$SU(2)_L\times SU(2)_R\times U(1)_{B-L}\times SU(3)_c$ 
have many attractive features. Parity would for example be unbroken 
at high energies and the known quarks and leptons fit very nicely 
and economically into fundamental representations of the
gauge group. In conventional left--right symmetric models
Higgses are however required to obtain the phenomenologically 
required symmetry breaking sequence:
$SU(2)_L\times SU(2)_R\times U(1)_{B-L}
\rightarrow SU(2)_L\times U(1)_{Y}
\rightarrow U(1)_{em}$.
In the light of the above considerations it is tempting to 
construct a left--right symmetric model without fundamental scalars, 
where the correct symmetry breaking sequence emerges dynamically.
A first condensate 
in the leptonic sector should break in a first step the left--right 
symmetry and a top condensate then the electro--weak symmetry. 
Both condensates together have to explain the correct mass to \GB
decay constant relations. This idea has been realized 
\cite{ALS1,ALS2} and in analogy to the BHL model all scalars 
are omitted in a first step from the left--right symmetric Lagrangian.
Acceptable four--fermi terms consistent with the symmetries are then
added and symmetry breaking by these terms can be studied. Such
a framework appears very attractive, since a Majorana condensate 
related to very high scales 
could occur in the neutrino sector and would not only provide the 
desired second condensate, but could also lead to a dynamical 
see--saw mechanism explaining the lightness of neutrinos. 
For the minimal fermion content of the model 
the auxiliary fields, which are naively square roots of the 
four--fermi terms, would contain a bi--doublet and triplets. 
This would also fit very nicely to the scenario which has been 
studied phenomenologically most intensively. Unfortunately it 
turns out that this simplest version does not work \cite{ALS1,ALS2}
since it is not possible to break 
parity in the required way. A way out is to postulate the 
existence of a new fermion which is a total gauge singlet \cite{ALS1,ALS2}. 
Then parity can break in the desired way, but the auxiliary fields 
contain now doublet scalars in addition to a bi--doublet. 
It is however interesting that this leads to another scenario with a 
see-saw mechanism. The reason is that the mass matrix of the 
neutral neutrino--like states is now a three by three matrix, 
with a huge entry in the 2--3 element. Upon diagonalisation one 
obtains the desired see--saw mechanism from doublets \cite{ALS1,ALS2}.

This left--right symmetric model is to our knowledge the first 
complete and successful attempt of a fully dynamical left--right 
symmetry breaking scenario. Due to the two condensates and the 
more involved relation between VEVs and fermion masses it does 
not produce a too high top mass like in the BHL case. Acceptable 
top mass values arise even for low values of $\Lambda$ eliminating 
thus the need for fine--tuning. The model leads to testable predictions 
in the Higgs mass spectrum. The lightest scalars are in this scenario 
naturally suppressed by one or two orders of magnitude compared to 
the typical left--right scale. If this left--right scale is not too 
high these light scalars could even show up at LEP II. 
Altogether this is nice toy model which has the desired \SM limit and 
which is therefore easily consistent with all data. If small 
deviations from the \SM 
were found one could lower some of the extra scalars and try to 
explain the effect. This has been demonstrated e.g. for the case 
of $R_b$ and other quantities of interest.

In summary we emphasised the fact that the current experimental 
situation may indirectly 
point towards the existence a scalar Higgs particle 
as unitarity partner of the \GBsp  We pointed out, that this would 
however not tell us anything about the fact whether this scalar 
is composite or fundamental. The current data do therefore not point 
to a solution of the hierarchy problem.
The question whether it is possible to build systematically 
models of dynamical symmetry breaking which lead to a composite 
Higgs and where effects beyond the \SM decouple emerged from these
considerations. We pointed out, that 
the so--called BHL model of electro--weak symmetry breaking has all 
desirable features. The BHL model is however unacceptable since it 
can not accommodate the correct top mass. We argued with the help 
of the Pagels--Stokar relation that this has systematic reasons and 
that the experimental top mass value is in general too small for a 
scenario with just a top condensate. 
We postulated therefore the existence of a sequential breaking 
of an extended gauge group with a second condensate. This idea 
has been realized in the class of left--right symmetric models,
but it is probably possible to build many other models which
have the desired features. Like the BHL model the left--right 
symmetric model has a \SM limit which allows it to hide behind 
the \SMp The left--right symmetric model can be made consistent 
with all data (including the top mass) even for low scales of new 
physics eliminating thus the need for fine--tuning. 
This illustrates that phenomenologically viable models can be 
built systematically. The scalar spectrum of the left--right
symmetric model was produced by NJL interactions. Like in the BHL 
case the required four--fermion terms could be Fierz 
rearranged to accommodate the model into a larger broken 
gauge group. 


\end{document}